\documentstyle[11pt]{article}

\begin{document}

\title{THE EMBEDDING OF SCHWARZSCHILD IN BRANEWORLD}

\author{EDMUNDO M. MONTE \thanks{edmundo@fisica.ufpb.br; edmundomonte@pq.cnpq.br}\\
Universidade Federal da Paraiba,  Departamento de Fisica,  58059-970}

\maketitle

\begin{abstract}
The braneworlds models were inspired partly by Kaluza-Klein's theory, where both the gravitational and the gauge fields are obtained from the geometry of a higher dimensional space. The positive aspects of these models consist in perspectives of modifications it could bring in to particle physics, such as: unification in a TeV scale, quantum gravity in this scale and deviation of Newton's law for small distances. One of the principles of these models is to suppose that all space-times can be embedded in a bulk of higher dimension. The main result in these notes is a theorem showing a mathematical inconsistency of the Randall-Sundrum braneworld model, namely that the Schwarzschild space-time cannot be embedded locally and isometrically in a five dimensional bulk with constant curvature,(for example AdS-5). From the point of view of semi-Riemannian geometry this last result represents a serious restriction to the Randall-Sundrum's braneworld model.
\end{abstract}

\section{Introduction}
\label{intro}
Recent alternative gravitacional theories involve the idea that the 3-spatial dimensions in which we live could be a 3-spatial-dimensional 'membrane' embedded in a much larger extra dimensional space, and that the hierarchy of interactions is generated by the geometry of the additional dimensions. Such ideas have led to extra dimensional theories which have verifiable consequences at the TeV scale. Since Newton's Law must be reproduced at large distances, gravity must behave at the milimeter scale as if there were only three spatial dimensions. Alternatively from point of view of General Relativity, since the gravitation must be reproduced at distances of solar system, where we have a space-time of Schwarzschild, gravity must behave as if there were only four dimensions (three spatial plus one time).\cite{Hewett},\cite{ADD},\cite{DGP}

Randall and Sundrum have proposed an interesting scenario of extra non-compact dimensions in which four-dimensional gravity emerges as a low energy effective theory, to solve the hierarchy problems of the fundamental interactions. This proposal is based on the assumption that ordinary matter and its gauge interactions are confined within a four dimensional hypersurface, the physical braneworld, embedded in a five-dimensional space and the fact that a bound state of a five-dimensional graviton exists and is localized near the physical braneworld.\cite{RS2},\cite{Rubakov}

The emphasis in the development of higher dimensional theories has recently shifted toward  the braneworld picture. A braneworld may be regarded as a space-time locally embedded in a higher dimensional space, the bulk,  solution of higher dimensional Einstein's equations. Furthermore, the embedded geometry  is assumed to  exhibit quantum  fluctuations with respect to  the extra dimensions at the  TeV scale of  energies. Finally, all  gauge interactions belonging to the standard model must remain confined to the four-dimensional space-time. Contrasting with  other  higher dimensional theories, the extra dimensions may be large and even infinite, with the possibility of being observed by  TeV accelerators. The embedding  conditions relate the  bulk  geometry  to the braneworld geometry, as it is clear from the  Gauss-Codazzi-Ricci equations.\cite{ADD},\cite{ME},\cite{Maeda} 

The Randall-Sundrum model consists of a three-brane embedded in a five-dimensional space which is asymptotically anti-de Sitter. A generic form of the metric in this space is
\[
ds^2=e^{-2\kappa|y|}{\bar g}_{\mu\nu}dx^\mu dx^\nu+dy^2,
\]
the four-dimensional metric ${\bar g}_{\mu\nu}$ being asymptotically flat. 
The asymptotic metric with ${\bar g}_{\mu\nu}
=\eta_{\mu\nu}$ satisfies the vacuum 
Einstein equations
\[
R_{mn}-{1\over 2}Rg_{mn}-\Lambda g_{mn}=0
\]
with cosmological constant $\Lambda=-6\kappa^2$, where $\kappa$ describes the curvature scale, which is assumed to be of order the Planck scale. Here we adopt the convention that the Greek indices take values 1-4, the Latin indices 1-5. The five dimensional anti-de-Sitter space is a space of constant curvature that depends of constant $\kappa$. We may assume without loss of generality that the constant is $K_{o}$.\cite{Giannakis-Ren}

In order to describe the real world, the Randall-Sundrum scenario has to satisfy all the existing tests of general relativity, based in the motion of material particles within the Schwarzschild braneworld, that it is given by four-dimensional geodesic equation.\cite{RS2},\cite{Duck},\cite{GT}

The purpose of this paper is to show a mathematical inconsistency of the Randall-Sundrum's braneworld model, namely, the Schwarzschild space-time seen as a braneworld cannot be embedded locally and isometrically in the AdS-5 bulk. 
 
\section{Embedding of Schwarzschild in a Bulk-5D of Constant Curvature}
\label{sec:1}
Now we prove that a class of Ricci flat space-times, as example the Schwarzschild space, cannot be embedded in the five dimensional space with constant curvature.\\

\vspace{0,5cm}
{\bf Theorem:}

\vspace{0,5cm}
{\it The Schwarzschild's space-time, with usual metric $g_{\mu\nu}$, can not be embedded locally and isometrically in the five dimensional space anti-de Sitter, where the curvature is $K_{o}\neq 0.$}\\
\vspace{0,5cm}

{\bf Proof:}
\vspace{0,5cm}

The Schwarzschild's  solution represents the  empty
space-time with spherical symmetry  outside of a body with spherical
mass $M$. Using spherical coordinates $(t,r,\theta ,\phi )$ this solution 
is given by $g$,
\[
\left( g_{\mu\nu}\right)=
\left( 
\begin{array}{cccc}
(1-2m/r) & 0 & 0 & 0 \\ 
0 & -(1-2m/r)^{-1} & 0 & 0 \\ 
0 & 0 & -r^{2} & 0 \\ 
0 & 0 & 0 & -r^{2}\sin ^{2}\theta 
\end{array}%
\right)
\]
where $ M =c^{2}mG^{-1}$, $m$ is an integration constant, $c$ is the speed of light and  $G$ is the
gravitational constant. We define the following regions:\cite{SchwTop}\\

a)The exterior Schwarzschild space-time  $(V_{4},g)$:\\

$ V_{4}=P_{I}^{2}\times S^{2}\;$; $\; P_{I}^{2}=\{(t,r)\in R^{2}/\; r>2m\}\;$\\

b) The Schwarzschild black hole  $(B_{4},g)$:\\

$ B_{4}=P_{II}^{2}\times S^{2}\;$; $\; P_{II}^{2}=\{(t,r)\in R^{2} /\; 0<r<2m\}\;$\\

In both cases, $S^{2}$  is the sphere of radius $r$ with metric $g$.

If we use the fundamental theorem for submanifolds: the four dimensional Schwarzschild space-time with usual metric $g$ can be embedded locally and isometrically in a five dimensional bulk of constant curvature $K_{o}$ if and only if there exists a symmetric tensor $b$ satisfying respectively the Gauss and Codazzi equations,\cite{Einsenhart},\cite{KN}
\begin{equation}
R_{\mu\nu\alpha\beta}=(b_{\mu\alpha}b_{\nu\beta}-b_{\mu\beta}b_{\nu\alpha})+K_{o}(g_{\mu\alpha}g_{\nu\beta}-g_{\mu\beta}g_{\nu\alpha})
\end{equation} and
\begin{equation}
b_{\mu\nu;\alpha}-b_{\mu\alpha;\nu}=0,
\end{equation}
where the (;) is the covariant derivative in relation to $g$ metric, $R_{\mu\nu\alpha\beta}$ are the components of curvature tensor of the Schwarzschild space-time and $b_{\mu\nu}$ are components of the extrinsic curvature in this space.

Contracting equation (1) and putting $R_{\nu\alpha}=0$ for Schwarzschild space-time,
\[
(b_{\mu\alpha}b_{\nu}^{\mu}-b_{\mu}^{\mu})=3K_{o}g_{\nu\alpha}.
\]
 
The tensor  $b_{\nu\alpha}$ must take one of the two canonical forms,
\[
(i) b_{\nu\alpha}=\lambda g_{\nu\alpha} \hspace{1,5cm} -K_{o}=\lambda^{2}
\]
\[
(ii) b_{\nu\alpha}=\lambda g_{\nu\alpha}+2\lambda u_{\nu}u_{\alpha}-2\lambda s_{\nu}s_{\alpha} \hspace{1,5cm} 3K_{o}=-\lambda^{2},
\]
where $u_{\nu}$, \hspace{0,1cm}$s_{\alpha}$ are vectors satisfying $-u_{\nu}u^{\nu}=s_{\nu}s^{\nu}=1$ and\hspace{0,1cm} $u_{\nu}s^{\nu}=0$.\cite{HK},\cite{Szekeres},\cite{Collinson}

Substituting $(i)$ into equation (1) results $R_{\mu\nu\alpha\beta}=0$ which contradicts the hypothesis that the Schwarzschild braneworld is not flat.
The expression in $(ii)$ can be written as: \cite{Collinson}

\begin{equation}
b_{\nu\alpha} = \sqrt{(+3K_{o})}g_{\nu\alpha}-\sqrt{(+12K_{o})}(l_{\nu}n_{\alpha}+n_{\alpha}l_{\nu})
\end{equation}

where the null vectors $l_{\nu}$ and $n_{\nu}$ are defined by 
\[
l_{\nu}=(s_{\nu}+u_{\nu})/\sqrt{2} \hspace{1,0cm}	and \hspace{1,0cm} n_{\nu}=(s_{\nu}-u_{\nu})/\sqrt{2}.
\]
Substituting (3) into (1),(2) and after of long calculus we then obtain an expression
\[
-4K_{o}l_{[\beta}g_{\mu][\alpha}l_{\nu]}=0. 
\] 	
Note that this expression also lead us to absurd, because $K_{o}\neq 0.$ $\triangle$

\section{Comments}
\label{sec:2}

We verify, with this result, that it is not possible to obtain the embedding of Schwarzschild space-time into a five dimensional bulk with constant curvature. Therefore, from point of view of embedding, there are geometric contraints which must be taken into account by braneworld models. Thus, we cannot have the anti-de-Sitter space as bulk for an important class of space-times, which is a serious restriction to Randall-Sundrum braneworld model. A Schwarzschild black hole cannot be embedded in the bulk AdS-5 bulk. It is important to take into consideration the fact that in the Randall-Sundrum model the braneworld examples acts as if they are fixed boundaries, which suggests that the embedding is rigid. 

The theorem proved above is relevant for experimental consequences of Randall-Sundrum's model, since some of  the experiments which are planned at the LHC imply in the generation of mini black holes from a TeV energy proton-proton collision within the mentioned braneworld model. According to the theorem, there are two possibilities. Either the black holes will not appear, because the model uses an AdS-5 bulk space. Or, on the other hand, the black holes will be observed and the braneworld model needs to be modified. It should be mentioned that we can obtain the embedding of Schwarzschild braneworld into a six dimensional pseudo-Euclidean bulk with different signatures, which allows changes in the topology of Schwarzschild brane\cite{SchwTop}.

\section{Acknowledgements}

I would like to thank Professor Marcos Maia and Professor Joel F. Neto for useful discussions. On the other hand I would also like to thank the FAPES-ES/CNPq  (PRONEX) and the FAPESQ-PB/CNPq (PRONEX) for their partial financial support.




\end{document}